\documentclass[aps,prd,twocolumn,showpacs,amsmath,amssymb,nofootinbib,eqsecnum, preprintnumbers]{revtex4-1}
 
\usepackage{hyperref} 
\usepackage{graphicx}
\usepackage{epstopdf}

 \newcommand{\sgn}{\mathop{\mathrm{sgn}}}

\newcommand{\be}{\begin{equation}}
\newcommand{\ee}{\end{equation}}
\newcommand{\ba}{\begin{eqnarray}}
\newcommand{\ea}{\end{eqnarray}}

\begin{document}
\title{Dynamical symmetries of the Kepler problem}

\author{Marco Cariglia}
\email{marco@iceb.ufop.br}
\affiliation{Universidade Federal de Ouro Preto, ICEB, Departamento de F\'isica.
  Campus Morro do Cruzeiro, Morro do Cruzeiro, 35400-000 - Ouro Preto, MG - Brasil}
 
\author{Eduardo Silva Ara\'ujo} 
\email{duduktamg@hotmail.com} 
\affiliation{Universidade Federal de Ouro Preto, ICEB, Departamento de F\'isica.
  Campus Morro do Cruzeiro, Morro do Cruzeiro, 35400-000 - Ouro Preto, MG - Brasil}

\date{\today}  

\begin{abstract} 
This work originates from a first year undergraduate research project on hidden symmetries of the dynamics for classical Hamiltonian systems, under the program 'Jovens talentos para a Ci\^encia' of Brazilian funding agency Capes. For pedagogical reasons the main subject chosen was Kepler's problem of motion under a central potential, since it is a completely solved system. It is well known that for this problem the group of dynamical symmetries  is strictly larger than the isometry group $O(3)$, the extra symmetries corresponding to hidden symmetries of the dynamics. By taking the point of view of examining the group action of the dynamical symmetries  on the allowed trajectories, it is possible to teach in the same project basic elements of as many important subjects in physics as: Hamiltonian formalism, hidden symmetries, integrable systems, group theory, and the use of manifolds. 
\end{abstract}

\maketitle

\section{Introduction} 
The Keplerian motion is a well  understood subject and a good platform to introduce students of Physics to concrete, real world problems. Depending on the level at which it is taught it may involve a direct study of the properties of the allowed trajectories, ellipses, parabolae and hyperbolae, a study of Hamiltonian mechanics, and a discussion of the role of dynamical symmetries generated by the angular momentum, associated to isometries, and by the Runge-Lenz vector, associated to hidden symmetries. The role of dynamical symmetries tends to be left to more advanced treatments since it is usually expressed in terms of the abstract Poisson brackets algebra of the conserved quantities of motion. 
 
The present work takes its motivation from a first year undergraduate research project under the program 'Jovens talentos para a Ci\^encia', sponsored by the Brazilian funding agency Capes \cite{Capes}. The aim is to introduce the student to the concept of dynamical and hidden symmetries using a concrete, well understood example, and introducing a point of view on the subject that does not require mastery of symplectic space methods. The project is at the same time designed to expose the student to symplectic methods as well as other important topics in physics such as integrable systems, group theory, use of manifolds. It should be noted however that the point of view taken, that the dynamical symmetry group of the Kepler problem induces a group action on allowed trajectories seen a single object, provides a simple way to explain what the dynamical symmetry group is and to geometrically visualise the action of hidden symmetries. 
 
The work is structured as follows. In sec\ref{sec:symplectic_geometry} we formulate Kepler's problem using Hamiltonian dynamics and basic symplectic geometry, introducing the Poisson brackets. In sec.\ref{sec:conserved_quantities} we discuss the conserved quantities of Keplerian motion: the angular momentum and the Runge-Lenz vector, and the concept of dynamical symmetry group that they generate. We comment on the specific property of the system of being maximally super-integrable and discuss the general concept of integrability and super-integrability. In sec.\ref{sec:allowed_trajectories} we discuss the allowed trajectories, ellipses, parabolae and hyperbolae, and the fact that there is a group action of the dynamical symmetry group that transforms trajectories into trajectories of the same energy. Then we calculate explicitly the non-trivial hidden symmetry action originated by the Runge-Lenz vector, both in infinitesimal and in finite form. The result is a smooth change in eccentricity of the conics performed at fixed energy that runs through all the alllowed values of the eccentricity. The action is thus transitive on trajectories of a given energy. Finally, in sec.\ref{sec:global_properties} we discuss the global properties of the manifold of allowed trajectories, and how the action of the dynamical symmetry group can be used to classify them. We finish in sec.\ref{sec:conclusions} with a summary and concluding remarks.


\section{Kepler's problem}
\subsection{Equations of motion, Hamiltonian dynamics and symplectic geometry\label{sec:symplectic_geometry}}  
Kepler's problem consists of the classical motion of a point particle with mass $m$ in flat three-dimensional space $\mathbb{R}^3$, under a central potential $V = - \frac{k}{r}$. Here $k$ is a positive constant and $r = \sqrt{x^2 + y^2 + z^2}$ is the distance from the origin, $x,y,z$ being cartesian coordinates. This finds applications for example in the study of the gravitational interaction of planets and comets with the Sun in the solar system, or in the attractive interactions of two electric charges of different sign. 
 
We will use the following notation: $\{ \hat{e}_x,\hat{e}_y,\hat{e}_z \}$ is an orthonormal cartesian base for vectors, $\vec{r} = x\hat{e}_x + y \hat{e}_y + z\hat{e}_z$ is the radial vector and $\hat{e}_r = \frac{\vec{r}}{r}$. $t$ is the absolute time coordinate and we will write time derivatives of functions $f(t)$ as $\frac{df}{dt} = \dot{f}$, $\frac{d^2f}{dt^2} = \ddot{f}$. The modulus of a vector $\vec{v}$ is indicated with $v$. The motion under a central  potential $V = - \frac{k}{r}$ is a well understood dynamical system and there are many excellent texts, both older and more recent, discussing its main properties, along with the properties of Hamiltonian systems and symplectic spaces in general. We refer the reader for example to \cite{Goldstein,Tong}, these two not representing an exhaustive list. 
 
Newton's equations of motion are 
\be \label{eq:Newton}
m \ddot{\vec{r}} = - \frac{k}{r^2} \hat{e}_r \, , 
\ee 
where the right hand side of the equation corresponds to the attractive central force. 
 
Eq. \eqref{eq:Newton} can be conveniently reformulated in symplectic space, that is the space defined by 6 coordinates $y^a$, $a=1,\dots, 6$, where $y^i = x,y,z$ for $i=1,2,3$, and $y^{i+3} = p_x, p_y, p_z$. $\vec{p}$ has the interpretation of a physical momentum vector. In symplectic space we can exchange eq.\eqref{eq:Newton}, that is a system of 3 second order equations, with the following system of 6 first order equations: 
\ba 
\dot{\vec{r}} &=& \frac{\vec{p}}{m} \, , \label{eq:Hamilton1} \\ 
\dot{\vec{p}} &=& - \frac{k}{r^2} \hat{e}_r \, . \label{eq:Hamilton2}
\ea 
These equations in turn can be given a geometrical origin. First define the Hamiltonian function in symplectic space as 
\be 
H (\vec{r}, \vec{p}) = \frac{p^2}{2m} - \frac{k}{r} \, , 
\ee 
physically representing the total energy written as kinetic plus potential energy. Then define the following antisymmetric matrix: 
\be \label{eq:inverse_symplectic}
\omega^{ab} = \sum_{i=1}^3 \left(  \delta^a_i \delta^b_{i+3} - \delta^a_{i+3} \delta^b_i \right) \, . 
\ee 
This is the so-called \textit{inverse symplectic matrix}. With this matrix we can define two operations. The first operation is the \textit{symplectic gradient} of a function $f(\vec{r}, \vec{p})$ as a 6-dimensional vector $X_f$ with components 
\be 
X_f^a = \sum_{b = 1}^6 \omega^{ab} \frac{\partial f}{\partial y^b} \, . 
\ee 
Using the symplectic gradient we can re-write eqs.\eqref{eq:Hamilton1}, \eqref{eq:Hamilton2} as 
\be \label{eq:eom_symplectic_gradient}
\dot{y}^a = X_H^a \, , 
\ee 
which has the interpretation of defining a trajectory $t \mapsto y^a(t)$ in symplectic space that is tangent to $X_H$. This trajectory is the \textit{Hamiltonian flux} passing through an appropriate initial point. 
 
The second operation we can define is a bracket, or multiplication, acting on the space of functions $f(y)$, $g(y)$ defined on symplectic space. We define it as 
\be 
\{ f, g \} = \sum_{a,b = 1}^6 \omega^{ab}  \frac{\partial f}{\partial y^a}  \frac{\partial g}{\partial y^b} \in \mathbb{R} \, . 
\ee 
This is called \textit{Poisson bracket}. The Poisson bracket is antisymmetric, $\{ f, g \} = - \{g, f\}$,  and it can be shown that is satisfies the \textit{Jacobi identity} 
\be 
\left\{ f, \left\{ g, h \right\} \right\} + \left\{ g, \left\{ h, f \right\} \right\} + \left\{ h, \left\{ f, g \right\} \right\} = 0 \, . 
\ee 
Then the equations of motion eq.\eqref{eq:Hamilton1}, \eqref{eq:Hamilton2} can also be re-written as 
\be 
\dot{y}^a = \{ y^a , H \} \, .  
\ee 
From this equation we can interpret $H$ as generator of time translations, $dy^a =  \{ y^a , H \} dt$. The time derivative of any function $f(\vec{r}, \vec{p})$ along an allowed trajectory $t \mapsto \left( \vec{r}(t), \vec{p}(t) \right)$ obeying \eqref{eq:eom_symplectic_gradient} can then be written as 
\be 
\dot{f} = \sum_{a=1}^6 \frac{\partial f}{\partial y^a} \dot{y}^a =  \sum_{a, b=1}^6 \frac{\partial f}{\partial y^a} \omega^{ab} \frac{\partial H}{\partial y^b} = \{ f, H \} \, . 
\ee 
As a consequence $H$ is always conserved on trajectories. 
 
\subsection{Conserved quantities, dynamical symmetries and Integrable systems\label{sec:conserved_quantities}} 
An important quantity in Kepler's problem is the angular momentum of the particle with respect to the origin of coordinates: 
\be \label{eq:L_def}
\vec{L} = \vec{r} \times \vec{p} \, . 
\ee 
It can be easily checked applying the equations of motion \eqref{eq:Hamilton1}, \eqref{eq:Hamilton2} that the angular momentum is conserved, $\dot{\vec{L}} = 0$. 
 
In general, given any regular function $f(\vec{r}, \vec{p})$ on symplectic space it is possible to generate an infinitesimal transformation of the coordinates 
\be \label{eq:transformation_generic}
\left\{ \begin{array}{lcl} \delta x^i &=& \eta \{  x^i, f  \} \, , \\ 
			    \delta p_i &=& \eta \{  p_i , f \} \, , 
\end{array} \right. 
\ee 
where $\eta$ is an infinitesimal parameter. It is possible to show that the inverse symplectic matrix is preserved under the transformation, $\mathcal{L}(\omega^{ab}) = 0$, where $\mathcal{L}$ is the Lie derivative, and therefore this is an infinitesimal canonical transformation. In particular, when $f$ is a conserved quantity, $\{ f, H \} = 0$, this means the Hamiltonian function is conserved along the transformation, $\delta H  = 0$. 
 
The angular momentum, defined above, provides three such independent conserved quantities and therefore three independent infinitesimal transformations, namely $\forall i = 1,2,3$: 
\be \label{eq:transformation_L}
\left\{ \begin{array}{lcl} \delta_{L^i} \, x^j  &=& \eta \, \{  x^j, L^i  \} = - \eta \, \epsilon^{ijk} x^k \, , \\ 
			    \delta_{L^i} \, p_j &=& \eta \, \{ p_j, L^i  \} =  - \eta \, \epsilon^{ijk} p_k \, , 
\end{array} \right. 
\ee 
where $\epsilon^{ijk}$ is the totally antisymmetric Levi-Civita symbol with $\epsilon^{123} = 1$. One can immediately recognize in eq.\eqref{eq:transformation_L} the form of an infinitesimal rotation. In fact the $L^i$ quantities satisfy the Poisson algebra 
\be \label{eq:O(3)_algebra}
\{ L^i , L^j \} = \sum_{k = 1}^3 \epsilon^{ijk} L^k \, . 
\ee 
This is the Lie algebra of the group $O(3)$ of rotations in 3 dimensions, associated to infinitesimal transformations of the group. So for every pair $(\vec{r}, \vec{p})$ in symplectic space eq.\eqref{eq:transformation_L} represents an infinitesimal transformation of the group $O(3)$. Moving from infinitesimal to finite transformations one obtains a \textit{group action} of the group $O(3)$ on symplectic space. 
 
The angular momentum $\vec{L}$ and the energy $H$ are not the only conserved quantities for Kepler's problem. Another important vectorial conserved quantity is the Runge-Lenz vector 
\be \label{eq:A_def}
\vec{A} = \vec{p} \times \vec{L} - mk \, \hat{e}_r \, . 
\ee 
This is quadratic in the momenta, as opposed to $\vec{L}$ which is linear, and a derivation that $\dot{\vec{A}}=0$ can be found in references \cite{Goldstein,Tong}. $H$, $\vec{L}$ and $\vec{A}$ can be seen to form a closed algebra under Poisson brackets, given by \eqref{eq:O(3)_algebra} and 
\be \label{eq:LA_algebra}
\{ L^i , A^j \} =  \sum_{k = 1}^3 \epsilon^{ijk} A^k \, ,  
\ee 

\be \label{eq:AA_algebra}
\{ A^i , A^j \} = - 2 m H \sum_{k = 1}^3 \epsilon^{ijk} L^k \, . 
\ee 
Equations \eqref{eq:O(3)_algebra} and \eqref{eq:LA_algebra} have the interpretation that, under the infinitesimal $O(3)$ transformation \eqref{eq:transformation_L}, $\vec{L}$ and $\vec{A}$ transform as vectors in $\mathbb{R}^3$. Eq.\eqref{eq:AA_algebra} shows that the algebra is closed. If we restrict to solutions with zero energy, $H=0$, the right hand side of \eqref{eq:AA_algebra} is zero and the algebra of $\vec{L}$ and $\vec{A}$ is that of $O(3) \ltimes \mathbb{R}^3$. For solutions with $H = E \neq 0$ we can introduce the new quantity $\vec{B} = \frac{\vec{A}}{\sqrt{2m |E|}}$, which satisfies 
\be \label{eq:LB_algebra}
\{ L^i , B^j \} =  \sum_{k = 1}^3 \epsilon^{ijk} B^k \, ,  
\ee 

\be \label{eq:BB_algebra}
\{ B^i , B^j \} = - \sgn (E) \sum_{k = 1}^3 \epsilon^{ijk} L^k \, , 
\ee
 where $\sgn(x) = x / |x|$ for $x\neq 0$, $\sgn(0) = 0$ is the signum function. The algebra is that of $O(4)$ for $E<0$, and $O(1,3)$ for $E>0$. Both groups admit two Casimir operators, and it can be shown that they are given by 
\ba 
C_1 &=& B^2 - \sgn (H) L^2  \, , \label{eq:C1} \\ 
C_2 &=& \vec{L} \cdot \vec{B}  \, . \label{eq:C2} 
\ea 
One can check directly that $\{ C_{(1,2)} , L^i\} = 0 = \{ C_{(1,2)} , B^i\}$ using the Poisson brackets. Then, for the specific dynamics of Kepler's problem one can substitute eqs.\eqref{eq:L_def}, \eqref{eq:A_def} into \eqref{eq:C1}, \eqref{eq:C2} and obtain $C_1 = \frac{mk^2}{2|H|}$, $C_2 = 0$. For the case $H=0$ it is possible to use $C_1 = A^2 = m^2 k^2$ and  $C_2 = \vec{L} \cdot \vec{A}$. 
 
The set $\{ H, C_1, C_2 \}$ represents a maximal set of independent mutually commuting functions on phase space. In general, such set can have at most $n$ elements, where $n$ is the dimension of the position space, 3 in our case. A Hamiltonian system admitting a maximal set of independent mutually commuting functions on phase space is called \textit{Liouville integrable}: this means that there is a set natural "action-angle" variables using which the system becomes trivially solvable. In fact, the Kepler system is \textit{maximally super-integrable}, because it admits 5 independent conserved quantities of the dynamics, meaning that motion in the 6-dimensional symplectic space occurs on a 1-dimensional curve once the 5 independent conserved quantities are fixed. In general, in $n$ dimension the  maximum number admissible is $2n -1$, given that one of the $2n$ variables must always remain free in order to describe the time evolution. There are 5 such independent quantities because the 7 quantities $\{ H, \vec{L}, \vec{A} \}$ are all conserved, but they are not all independent, the functions $C_1$ and $C_2$ providing two constraints. 
 
The functions $\vec{L}$, $\vec{B}$ generate, for a fixed energy $H=E$, a group of transformations according to the rule \eqref{eq:transformation_generic}. Each of the transformations preserves the energy, $\delta H = 0$, and this is called the \textit{Dynamical Symmetry Group}. The $O(3)$ part of the dynamical symmetry group represents \textit{isometries}, i.e. rigid rotations of the 3-dimensional configuration space. It is possible to show that the remaining dynamical symmetries cannot be written as isometries. They are genuine transformations in symplectic space, not in configuration space, see for example \cite{MarcoDavidPavel2011} for a discussion of such symmetries for generic Hamiltonian systems, and are called \textit{Hidden Symmetries} of the dynamics. 
 
It is common practice to introduce the dynamical symmetry group algebraically as done in this section, i.e. presenting the generators of the group and their Poisson algebra. While correct and useful, this approach is abstract and tends to make it difficult to interpret the action of the hidden symmetry transformations. Some authors instead display explicitly the action of the dynamical symmetry group by extending the symplectic space to a bigger space, for Kepler's problem see for example \cite{Rogers}. This approach is interesting although it seems it may be necessary to search for the correct set of extended coordinates on an individual, problem by problem, basis. 
In the next sections we show how for the Kepler problem the dynamical symmetry group acts transforming trajectories into trajectories of the same energy. Therefore there is an explicit group action on the space of trajectories, and no extended space needs to be invoked. There already exist works in the literature showing compatible results, for example that hidden symmetries in the Kepler motion can change negative energy trajectories, that are ellipses, into ellipses of a different eccentricity \cite{Mostowski2010}. In this work we will analyse explicitly only the Kepler problem, and for this all types of trajectories, ellipses, hyperbolae and parabolae, because of the focused nature of the current research project. However, even if we will not discuss the general case, the underlying reasoning used here applies to generic systems. The generic case will be discussed in a separate work. 
 The hidden symmetry transformations become transformations that alter the shape of the trajectories, while isometries keep the shape unchanged. The action of the symmetry group is well defined on the whole manifold of the possible trajectories, and by taking the quotient with respect to the group action we can classify the different trajectories.

\subsection{Allowed trajectories and group actions\label{sec:allowed_trajectories}} 
It is a well known fact that the allowed trajectories that do not pass through the origin in Kepler's problem are ellipses, hyperbolae and parabolae, according to whether the total energy $E$ is negative, positive or zero. There exist also arbitrary energy straight line solutions that correspond to a particle running into, or leaving from, the center of coordinates. 
 
We will focus on the conical trajectories, the straight line solutions will be recovered as a limiting case of the conics for $L \rightarrow 0$. The conical trajectories lie in a plane perpendicular to the vector $\vec{L}$.  In this plane we can introduce cartesian coordinates $x, y$ and 2-dimensional radial coordinates $\rho , \theta$, with $\rho = \sqrt{x^2 + y^2}$ the distance from the center and $\theta$ the angle with the $x$ axis. A generic conic with one focus at the origin of the coordinates can be parameterised as 
\be 
\frac{1}{\rho} = C \left[ 1 + e \cos\left(\theta - \theta^* \right) \right] \, , 
\ee 
where $e$ is the \textit{eccentricity} and $\theta^*$ the angle of the perihelion, that is the angle between the point of minimum $\rho$ and the $x$ axis. For $e<1$ this is an ellipse, in particular a circle when $e=0$, for $e=1$ a parabola and for $e>1$ a hyperbola. 
 
In the specific Kepler problem one has 
$C = \frac{mk}{L^2}$, and 
\be 
e = \sqrt{1+ \frac{2EL^2}{mk^2}} = \frac{A}{mk} \, . 
\ee 
In particular circles have $E = - \frac{mk^2}{2L^2}$. We are excluding for now the cases with $L=0$, which are the straight lines. The direction of the perihelion is given by that of $\vec{A}$. 
 
We can parameterise a possible trajectory using the following five variables: the components of $\vec{L}$ and $\vec{A}$, subject to the constraints that $C_1$ and $C_2$ in eqs.\eqref{eq:C1}, \eqref{eq:C2} are constant. The reason five coordinates are required is the following: one can think that for each point $(\vec{r}, \vec{p})$ in symplectic space there passes a unique trajectory satisfying eqs.\eqref{eq:Hamilton1}, \eqref{eq:Hamilton2}. This gives six coordinates, but one of these is redundant: the time coordinate along the trajectory. Similarly, trajectories of a given fixed energy $E$ are described by four coordinates. We can also choose another set of local coordinates. First, for each given orientation of $\vec{L}$ we define smoothly coordinates $x, y$ in the perpendicular plane. This can only be done locally, since it is equivalent to choosing a vector $\hat{e}_x$ perpendicular to $\hat{n} = \frac{\vec{L}}{L} \in S^2$, and by the hairy ball theorem the choice cannot be done at the same time globally and smoothly. However, as we will see in section \ref{sec:global_properties}, we can limit ourselves to consider only half of the sphere $S^2$ and therefore this limitation does not apply. So we  can define a $\theta^*$ variable as the angle between $\vec{A}$ and the chosen x axis. Then we can use as variables the energy $E$, the angular momentum $\vec{L}$ which can be decomposed into its modulus $L$ and the associated unit norm vector $\hat{n} = \frac{\vec{L}}{L}$, and the perihelion angle $\theta^*$. 
 
We now show that the dynamical symmetry group induces an action on the space of trajectories. We introduce the following notation. For a given function $f(\vec{r}, \vec{p})$ let $\Phi_{(f,s)}$ be a map of symplectic space into itself, such that $\Phi_{(f,s)}: (\vec{r},\vec{p}) \mapsto \Phi_{(f,s)} (\vec{r},\vec{p})$, which is the image of $(\vec{r},\vec{p})$ under the finite transformation generated by \eqref{eq:transformation_generic}, for a finite parameter $s$. If $f$ is a constant of motion, $\{f, H \} = 0$, then $\Phi_{(f,s)}$ is a finite canonical transformation and one element of the dynamical symmetry group. Since $f$ and $H$ Poisson commute then $\Phi_{(f,s)}$ commutes with the Hamiltonian flow $\Phi_{(H,t)}$ for any value of $t$ and $s$. Consider now a trajectory passing through $(\vec{r}_0,\vec{p}_0)$: this is given by all points of the kind $\Phi_{(H,t)}(\vec{r}_0,\vec{p}_0)$, for some allowed $t$. If we act with $\Phi_{(f,s)}$ on any of these points we get 
\be 
\Phi_{(f,s)}  \Big( \Phi_{(H,t)}(\vec{r}_0,\vec{p}_0) \Big) = \Phi_{(H,t)}  \Big( \Phi_{(f,s)}(\vec{r}_0,\vec{p}_0) \Big) \, , 
\ee 
showing that this is the trajectory associated to the initial point $ \Phi_{(f,s)}(\vec{r}_0,\vec{p}_0)$. Then this means that $\Phi_{(f,s)}$ transforms a trajectory into another trajectory, with the same energy. 
 
For the Kepler problem, the $O(3)$ transformations are straightforward, since they correspond to rigid rotations of trajectories where $\vec{L}$ and $\vec{A}$ rotate accordingly without changing modulus. We want now to examine the action of the hidden symmetries, the transformations generated by $\vec{A}$. We do this in the following way. We start with a reference system where $\vec{L}$ lies in the $z$ direction and $\vec{A}$ in the x direction. Then we apply an infinitesimal transformation according to \eqref{eq:transformation_generic} and infer the result of the corresponding finite transformation. For a transformation generated by $A_x$, using the algebra \eqref{eq:LA_algebra}, \eqref{eq:AA_algebra}, we find that the only non-zero transformation is  
\be 
\delta A_y = \eta \{ A_y, A_x \} =  \eta \, 2 m E L , 
\ee 
with $\delta L^2 = 0 = \delta A^2$. This is then the infinitesimal version of a rotation around the $\vec{L}$ axis, and belongs to the $O(3)$ isometry subgroup. Similarly, the transformation generated by $A_z$ has 
\be 
\delta L_y = \eta \{ L_y, A_z \} = - \eta A \, , 
\ee 
with $\delta L^2 = 0 = \delta A^2$, and is a rotation around the $\vec{A}$ axis. The only non-trivial transformation is the one give by $A_y$, which has 
\be \label{eq:delta_LA}
\left\{ \begin{array}{lcl} 
	\delta L_z &=& \eta \{ L_z, A_y \} = - \eta A \, , \\ 
	\delta A_x &=& \eta \{ A_x, A_y \} = - \eta \, 2mEL \, . 
\end{array} \right. 
\ee 
Thus the amplitude of both $\vec{L}$ and $\vec{A}$ change, respecting $\delta C_1 = 0 = \delta C_2$ as can be explicitly checked. A direct calculation shows that under this transformation 
\be \label{eq:delta_e}
\delta e = - \eta \frac{2 E L}{k} = - \eta \sgn(E) \sqrt{2m |E|} \sqrt{|e^2-1|} \, . 
\ee 
Then the corresponding finite transformation will be a change in eccentricity at fixed energy. For parabolae, that have $E=0$, the eccentricity stays constant and equal to $1$, and only the amplitude of $\vec{L}$ changes, which is consistent with the general condition $A^2 = m^2 k^2 + 2m E L^2$. Then changing $L$ amounts to parabolae with different distances between the perihelion and the origin of the coordinates. 

It is easy to find the finite form of transformations \eqref{eq:delta_LA}, \eqref{eq:delta_e}. It is convenient to use the identities $A = mk e$, $A^2 = m^2 k^2 + 2 m E L^2$, and $e = \sqrt{1 + \sgn(E) \tilde{L}^2}$, where we defined the variable $\tilde{L} = \sqrt{\frac{2|E|}{mk^2}} L$. In terms of these eqs.\eqref{eq:delta_LA}, \eqref{eq:delta_e} are integrated as follows. If $E<0$ then $0\le \tilde{L} \le 1$ and 
\ba 
\tilde{L}(s) &=&  \cos \left( \sqrt{2m|E|} s + \cos^{-1} \tilde{L}_0 \right) \, , \\ 
e(s) &=&  \sin \left( \sqrt{2m|E|} s + \sin^{-1} \tilde{e}_0 \right) \, , \label{eq:ellipses_eccentricity_change}
\ea 
where $s$ is the transformation finite parameter, $0 \le e \le 1$ and $e^2 + \tilde{L}^2 = 1$. It can be seen that this transformation deforms a circle, $e=0$, $\tilde{L}=1$, into ellipses of higher and higher eccentricity, with limit $e=1$, $L=0$, which is segment of straight line and not, in order to clarify possible doubts, a parabola. 
\begin{figure}[htb]
\includegraphics[width=0.5\linewidth]{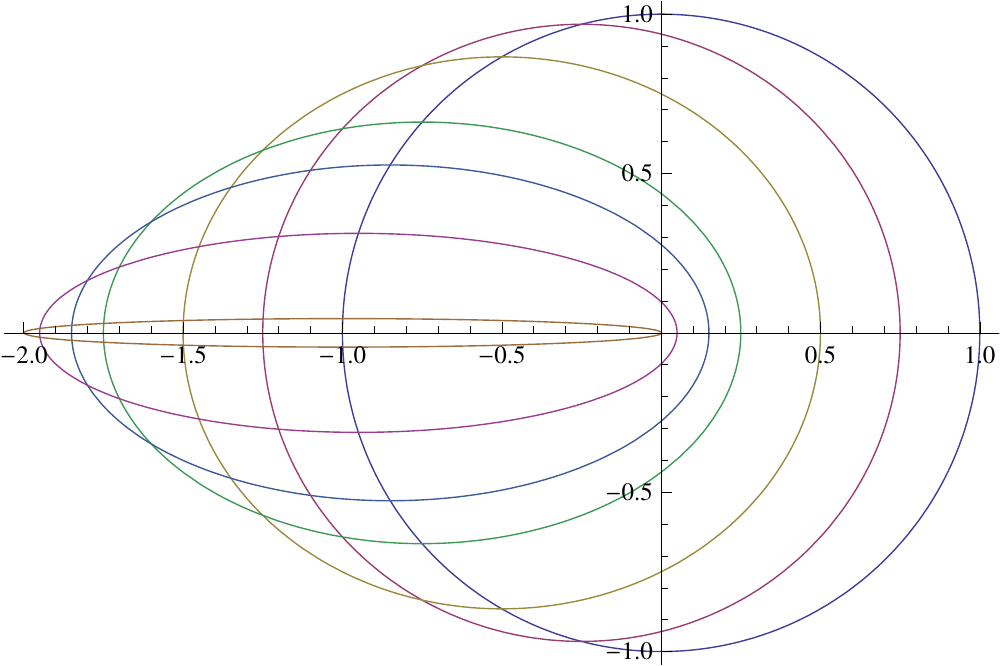}
\caption{\label{fig:ellipses}The finite transformation \eqref{eq:ellipses_eccentricity_change} is displayed for the following values of the eccentricity: 0, 0.25, 0.5, 0.75, 0.85, 0.95, 0.999. The ellipses lay in the $x$, $y$ plane, with foci on the $x$ axis. To generate the image we have set $\frac{k}{2|E|}=1$.}
\end{figure}

For $E=0$ the transformation is jut 
\ba \label{eq:parabolae_L_change}
L(s)  &=&  - mk s + L_0 \, , \\ 
e(s) &=&  e_0 \, , 
\ea
with no change in eccentricity - parabolae are mapped into parabolae, since the energy is unchanged - but with a varying $L$ which means that the shape of the parabola changes. For $L \rightarrow 0$ the conic degenerates into a straight line. 
\begin{figure}[htb]
\includegraphics[width=0.5\linewidth]{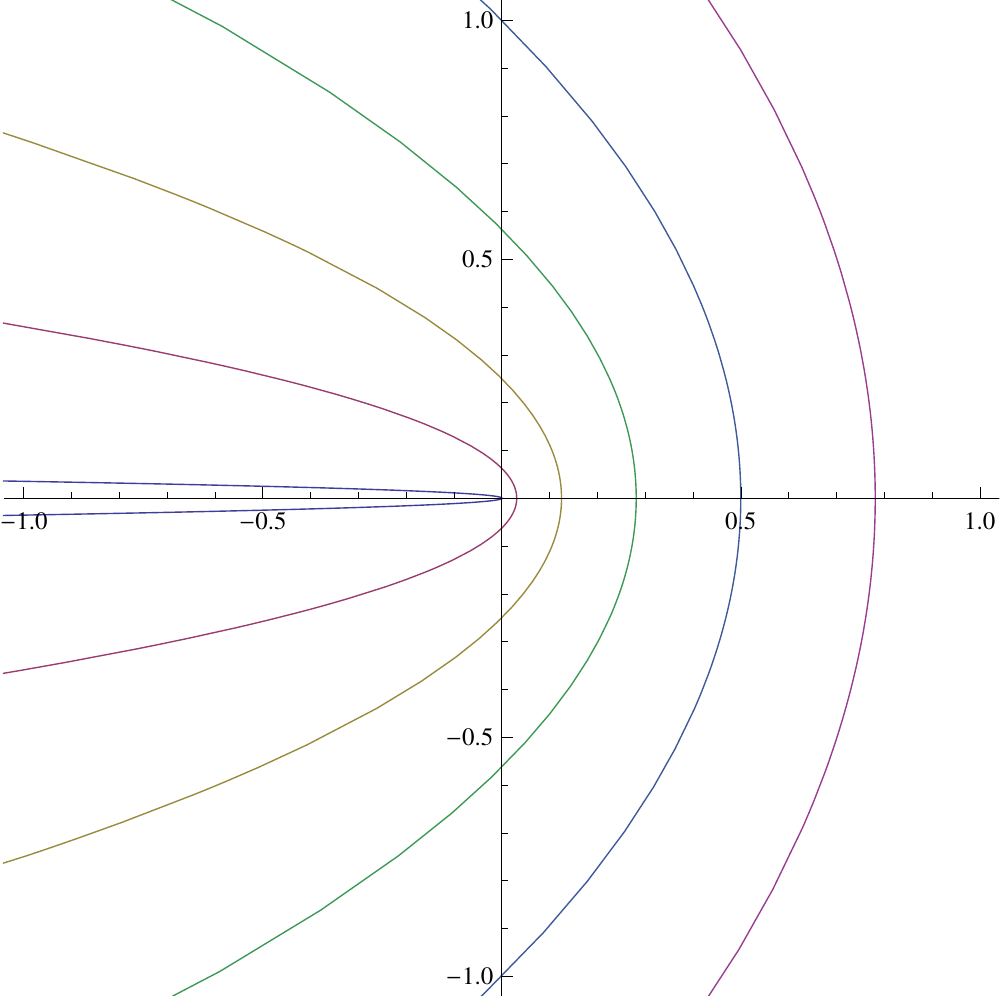}
\caption{\label{fig:parabolae}The finite transformation \eqref{eq:parabolae_L_change} is displayed for the following values of the parameter $L$: 0.025, 0.25, 0.5, 0.75, 1, 1.25. The parabolae lay in the $x$, $y$ plane, with focus on the $x$ axis. To generate the image we have set $mk = 1$.}
\end{figure}

Lastly, for $E>0$ the variable $\tilde{L}$ satisfies $\tilde{L} \ge 0$ the transformations are given by 
\ba 
\tilde{L} &=&  \sinh \left( - \sqrt{2m|E|} s + \sinh^{-1} \tilde{L}_0 \right) \, , \\ 
e &=&  \cosh \left( - \sqrt{2m|E|} s + \cosh^{-1} \tilde{e}_0 \right) \, , \label{eq:hyperbolae_eccentricity_change}
\ea 
with $e^2 - \tilde{L}^2 = 1$, and the hyperbola degenerating into a straight line for $\tilde{L} \rightarrow 0$. 
\begin{figure}[htb]
\includegraphics[width=0.5\linewidth]{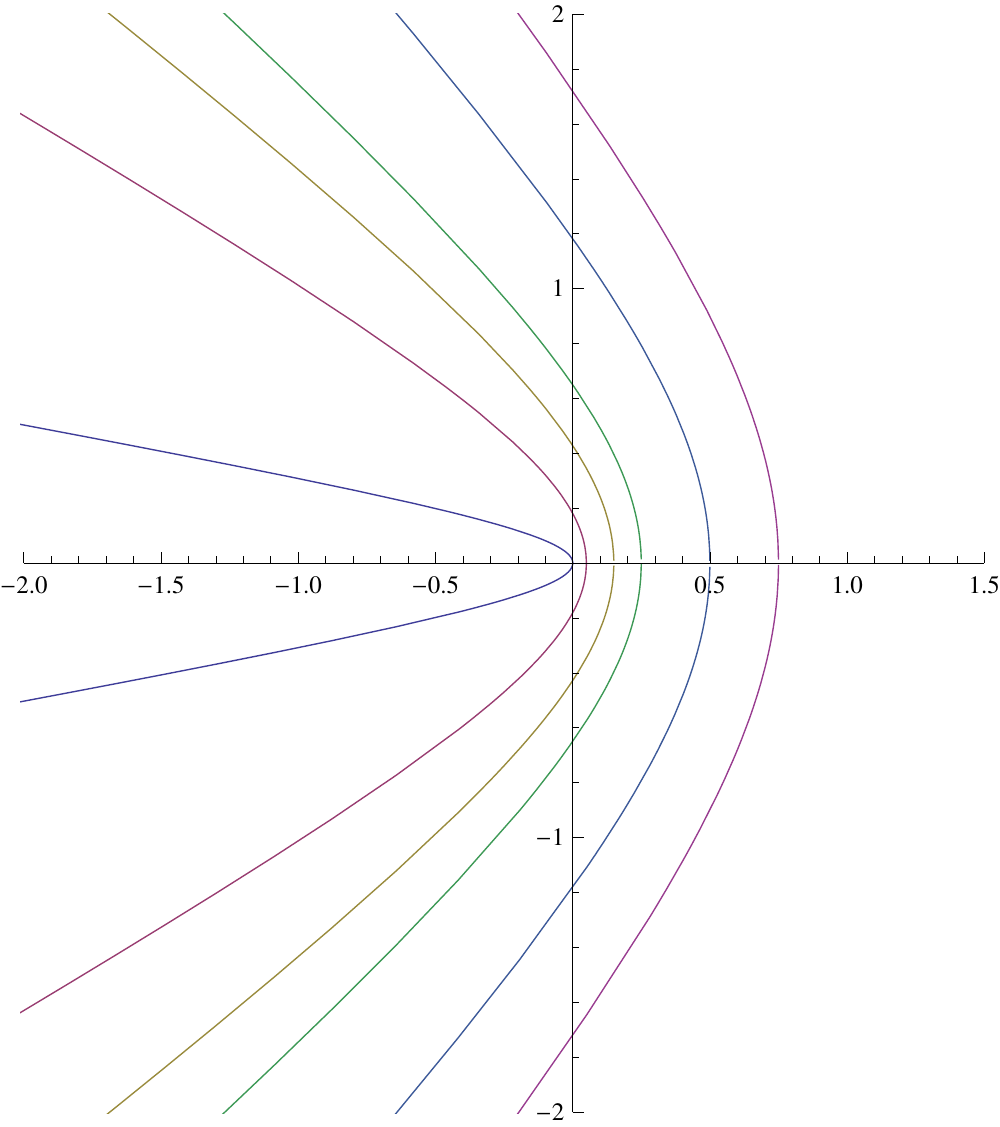}
\caption{\label{fig:hyperbolae}The finite transformation \eqref{eq:hyperbolae_eccentricity_change} is displayed for the following values of the eccentricity: 1.0005 , 1.05, 1.15, 1.25, 1.5 and 1.75. The hyperbolae lay in the $x$, $y$ plane, with foci on the $x$ axis. To generate the image we have set $\frac{k}{2|E|}=1$.}
\end{figure}

\subsection{The space of allowed trajectories as a manifold with global properties\label{sec:global_properties}} 
 One of the advantages of thinking of the dynamical symmetry group as a group acting on the space of allowed trajectories is that it allows to consider the latter space as a global object on its own, and to classify it according to the group action. 
 
We can start by studying separately the sections of constant $E$. For $E<0$ we have ellipses with $0 \le \tilde{L} \le 1$. $\tilde{L} = 1$ represents the circles, or equivalently $E = - \frac{mk^2}{2 L^2}$. To describe a general ellipse we need to specify 3 coordinates for $\vec{L}$, which amounts to specifying the orbital plane and the value of $L$, and a fourth coordinate corresponding locally to $\theta^*$ or in general to $\vec{A}$ subject to the constraints $C_1$ and $C_2$. This last freedom can be parameterised by a variable on the circle $S^1$. Given that $L \le \sqrt{\frac{mk^2}{2|E|}}$ one might think that the variable $\vec{L}$ should lie within a sphere of radius $\sqrt{\frac{mk^2}{2|E|}}$, which increases to infinity when the energy increases to zero. However, the operation $\vec{L} \rightarrow - \vec{L}$ is redundant since it yields the same orbital plane and the same value of $L$, and should be quotiented out. So $\vec{L}$ lies in a closed ball of radius $\sqrt{\frac{mk^2}{2|E|}}$ modulo the inversion operation, $\vec{L} \in S \left(\sqrt{\frac{mk^2}{2|E|}} \right) / \mathbb{Z}_2$. For all values of $\vec{L}$ inside the ball the eccentricity is different from zero and there is an $S^1$ freedom to rotate the ellipse around the $\vec{L}$ axis. However, for $\vec{L}$ on the surface of the ball the trajectories are circles and the $S^1$ freedom disappears: rotations around the $\vec{L}$ axis no longer generating new trajectories, so the $S^1$ collapses into a point. We can therefore picture the space of trajectories of a given negative energy $E$ as a cone with base given by $S \left(\sqrt{\frac{mk^2}{2|E|}} \right) / \mathbb{Z}_2$, the tip of the cone lying on the border of the sphere. Since using the dynamical symmetries we can change the direction of $\vec{L}$ and $\vec{A}$ (isometries), and their moduli (the non-trivial hidden symmetry) then the action of $O(4)$ is transitive on trajectories of negative energy. In other words, if we use dynamical symmetry transformations as an equivalence relation on the space of trajectories of a given negative energy $E<0$, then there is only one representative for each value of the energy. 
 
For $E\ge 0$ the analysis is similar but the $S^1$ circle does not collapse at any point, and moreover $\tilde{L}$ is unbounded from above, so the fixed energy manifold is given by $\mathbb{R}^3 / \mathbb{Z}_2 \times S^1$. In both cases the action of $O(4)$ is transitive and there is only one representative per given value of the energy.

\section{Conclusions\label{sec:conclusions}} 
We have discussed the classical Kepler problem from the point of view of its dynamical symmetries. While it is well known that the group of symmetries of the dynamics is strictly larger than $O(3)$, this is normally discussed in terms of Poisson brackets algebra and infinitesimal canonical transformations. We have taken here the point of view that the group of symmetries of dynamics acts on the space of allowed trajectories as a whole, and found the explicit form of the finite group transformations. We have shown that the dynamical symmetry group acts transitively on the space of trajectories of fixed energy $E$, for any allowed value of the energy, and we discussed the global structure of the manifold of allowed trajectories. Thus in a single project it has been possible to touch on several important subjects in physics.

\vspace{0.2cm}

\section*{Acknowledgments}
E. Silva Ara\'ujo acknowledges financial support from Capes, under the program 'Jovens talentos para a Ci\^encia'. 



\providecommand{\href}[2]{#2}\begingroup\raggedright\endgroup

\end{document}